\pdfoutput=1
\documentclass[sigconf,nonacm]{acmart}
\AtBeginDocument{%
  \providecommand\BibTeX{{%
    \normalfont B\kern-0.5em{\scshape i\kern-0.25em b}\kern-0.8em\TeX}}}

\setcopyright{acmcopyright}
\copyrightyear{2023} 
\acmYear{2023} 
\setcopyright{acmcopyright}\acmConference[SIGCSE 2023] {Proceedings of the 54th ACM Technical Symposium on Computer Science Education V. 1}{March 15--18, 2023}{Toronto, ON, Canada.}
\acmBooktitle{Proceedings of the 54th ACM Technical Symposium on Computer Science Education V. 1 (SIGCSE 2023), March 15--18, 2023, Toronto, ON, Canada}
\acmPrice{15.00}
\acmDOI{10.1145/3545945.3569882}
\acmISBN{978-1-4503-9431-4/23/03}

\usepackage{multirow}
\usepackage{balance}
\begin{document}

%%
%% The "title" command has an optional parameter,
%% allowing the author to define a "short title" to be used in page headers.
% \title[Identifying Different Types of Students in Functional Programming Assignments]{Identifying Different Student Clusters in Functional Programming Assignments: from Quick Learners to Struggling Students}

\title[Identifying Different Types of Students in Functional Programming Assignments]{Identifying Different Student Clusters in Functional Programming Assignments: From Quick Learners to Struggling Students}

%%
%% The "author" command and its associated commands are used to define
%% the authors and their affiliations.
%% Of note is the shared affiliation of the first two authors, and the
%% "authornote" and "authornotemark" commands
%% used to denote shared contribution to the research.

%%
%% By default, the full list of authors will be used in the page
%% headers. Often, this list is too long, and will overlap
%% other information printed in the page headers. This command allows
%% the author to define a more concise list
%% of authors' names for this purpose.

%%
%% The abstract is a short summary of the work to be presented in the
%% article.

% Functional programming and 
% To switch to developing in a pure functional style, they have to make a transition in their thinking and their approach to development.

\begin{abstract} 
Instructors and students alike are often focused on the grade in programming assignments as a key measure of how well a student is mastering the material and whether a student is struggling. This can be, however, misleading. Especially when students have access to auto-graders, their grades  may be heavily skewed. 

In this paper, we analyze student assignment submission data collected
from a functional programming course taught at McGill university incorporating a wide range of features. In addition to the grade, we
consider activity time data, time spent, and the number of
static errors. This allows us to identify four  clusters of students:
"Quick-learning", "Hardworking", "Satisficing", and "Struggling"
through cluster algorithms. We then analyze how work habits, working
duration, the range of errors, and the ability to fix errors impact different clusters of students. This structured analysis provides valuable insights for instructors to actively help different types of students  and emphasize different aspects of their overall course design. It also provides insights for students themselves to understand which aspects they still struggle with and allows them to seek clarification and adjust their work habits. 
\end{abstract}

%%
%% The code below is generated by the tool at http://dl.acm.org/ccs.cfm.
%% Please copy and paste the code instead of the example below.
%%
% \begin{CCSXML}
% <ccs2012>
% <concept>
% <concept_id>10003456.10003457.10003527.10003540</concept_id>
% <concept_desc>Social and professional topics~Student assessment</concept_desc>
% <concept_significance>500</concept_significance>
% </concept>
% </ccs2012>
% \end{CCSXML}

% \ccsdesc[500]{Social and professional topics~Student assessment}


% \ccsdesc[500]{Social and professional topics~Computer science education}

% \author{Chuqin Geng, Wenwen Xu, Yingjie Xu, Brigitte Pientka and Xujie Si}
% \affiliation{%
%   {\tt \{chuqin.geng, xiaojie.xu, haolin.ye\}@mail.mcgill.ca, {bpientka, xsi}@cs.mcgill.ca}
%   \institution{McGill University}
% }

\author{Chuqin Geng}
\affiliation{%
  \institution{McGill University}
  \country{Montreal, QC, Canada}}
\email{chuqin.geng@mail.mcgill.ca}

\author{Wenwen Xu}
\affiliation{%
  \institution{McGill University}
  \country{Montreal, QC, Canada}}
\email{wenwen.xu2@mail.mcgill.ca}

\author{Yingjie Xu}
\affiliation{%
  \institution{McGill University}
  \country{Montreal, QC, Canada}}
\email{yj.xu@mail.mcgill.ca}

\author{Brigitte Pientka}
\affiliation{%
  \institution{McGill University}
  \country{Montreal, QC, Canada}}
\email{bpientka@cs.mcgill.ca}

\author{Xujie Si}
\affiliation{%
  \institution{McGill University}
  \country{Montreal, QC, Canada}}
\email{	xsi@cs.mcgill.ca}

%%
%% Keywords. The author(s) should pick words that accurately describe
%% the work being presented. Separate the keywords with commas.

%% A "teaser" image appears between the author and affiliation
%% information and the body of the document, and typically spans the
%% page.

%%
%% This command processes the author and affiliation and title
%% information and builds the first part of the formatted document.
\maketitle
\pagestyle{empty}

\section{Introduction}
Online programming environments, such as RoboProf ~\cite{10.1145/305786.305904} for C++, DrScheme 
~\cite{10.1145/284563.284566, 10.1017/S0956796801004208} for Scheme or, more recently, Mumuki ~\cite{10.1145/3159450.3159579}
, offer immense potential to enhance the students' educational experience in large-scale programming-oriented courses. They not only lower the entry barrier for beginners but often feature auto-grading facilities that allow students to run and get feedback on their code while they are developing their programs, giving them the opportunity to fix bugs and address errors in their understanding right away. While having access to immediate feedback on their code has been recognized to significantly improve student learning outcomes and engagement (see, e.g., ~\cite{sherman2013impact,wilcox2015role,gramoli2016mining}),
instructors and students alike are often too focused on the grade as a key measure of competency. Especially when students have access
to auto-graders, the students' grades may be heavily skewed and misleading. %\textcolor{red}{shorten the auto-grader part}

This paper develops a data-driven approach to better
understand students' behavior when solving programming assignments in
a functional programming course. In addition to the grade, we propose
to consider additional factors such as the number of static
errors and total time spent on solving programming assignments to
identify student clusters. 
%
% the number and kinds of static errors, total time spent on the assignment, activity time data, and work habits. 
Using this methodology, we analyze the assignment submission data
collected in a functional programming course taught at McGill university which uses the Learn-OCaml online programming platform
~\cite{canou16,Canou_2017,hameer2019}.  
This allows us to identify four student clusters: \emph{"Quick-learning"}, \emph{"Hardworking"},
\emph{"Satisficing"}, and \emph{"Struggling"}. While the first two clusters can be characterized as maximizers, i.e. students strive to achieve the highest possible grades and continue to improve their work, they still differ in the amount of time and effort spent on completing a given homework. In contrast, satisficing\footnote{The term ``satisficing'' was introduced by H. Simon \cite{simon:PR56}
to describe a decision-making process in which an individual makes a choice that is satisfactory rather than optimal. 
%It morphs the verb ``satisfy'' and ``suffice''.
}
  students accept a
possibly non-optimal outcome as ''good enough'' 
allowing them to adequately achieve their goals by saving time and effort.
% Struggling students  are 
% exhibit similar challenges as the \emph{Harworking} students, but do not manage to overcome them.
We further analyze these clusters with respect to work habits and the number and kinds of errors that
are prevalent. This leads to four key insights: 

\begin{itemize}
    \item Leveraging the notion of \emph{chronotype} - a circadian
      typology in humans and animals, we confirm that a work pattern where students tend to work in the morning is related to academic success. In particular, quick learners tend to work more in the morning, while other clusters of students rely more on afternoons and evenings. 

    \item In general, starting on the homework early is related to higher grades. However, we also noticed that satisficing students start relatively late but finish the earliest. This further emphasizes that satisficing students aim for satisfactory results rather than the optimal one. At the same time, satisficing students have one of the lowest numbers of programming errors suggesting that they struggle significantly less with static errors than for example hardworking students. % This seems to indicate that they still grasp functional programming concepts.

%aims for a satisfactory or adequate result, rather than the optimal solution. Instead of putting maximum exertion toward attaining the ideal outcome, satisficing focuses on pragmatic effort when confronted with tasks

    \item Our analysis of static errors shows that syntax and type
      errors are prevalent among all students. Further, students
      continue to struggle with these errors throughout the
      semester. In addition, our analysis points to other common
      mistakes such as non-exhaustive case analysis and the use of unbound variables. 
% This seems to suggest that instructors could and possibly should provide a better way to avoid

   \item Taking into account students' ability to fix static errors,
     i.e. how many tries a student needs to fix a particular error, we
     notice that the failure/success ratio is particularly high for
     hardworking students. This highlights both their desire and drive
     to get the best possible grade, but also that their path is full
     of small stumbling blocks.
% which they do not understand how to avoid.

% \item Using a fine-grained analysis of the kinds of static errors
%       students encounter, we identify areas where students
%       struggle. While, general static errors such as type and syntax
%       errors are dominant, our data also suggests that students
%       struggle with concepts specific to typed functional programming
%       and recursive programming such as the use of first-class
%       functions and pattern matching. We also investigate the number
%       and kinds of errors that are prevalent in each the student
%       clusters. This suggests that \emph{Hardworking} students
%       struggle significantly more in all error categories than
%       \emph{quick learners}. In fact they struggle more than \emph{satisficing} students despite the fact that their grade is higher. \textcolor{red}{Too long}

\end{itemize}

We believe our proposed set of features and data-driven analysis can provide instructors with a clearer and more
detailed picture of students' behaviours and performance. This in turn may be used to adjust how some concepts, such as how to avoid static errors, are taught. It may also be used to design different strategies
for different students to enhance the students' learning experience. Furthermore, this data may be interesting to students themselves to better understand how well they do in a class and identify areas where they can actively make changes and seek help early.

\section{Related Work}\label{sec:relatedWork}

Analyzing student data in programming courses is a central topic in learning analytics, and it is gaining increasing attention with the recent advances in storing and processing data. One of the core aims of analyzing student data is to understand student behaviours, and in turn, improve student learning experience \cite{hundhausen_olivares_carter_2017}.

% Piech et al. \cite{piech2012modeling} uses clustering algorithms along with other machine learning techniques to autonomously create a graphical model of how students progress in an introductory programming course. Their clustering algorithm  {\color{red}{XXX as a feature}}.

Over the past decade, there have been several studies that focus on
identifying groups of students using cluster analysis. For example, Emerson et al. \cite{emerson_smith_rodriguez_wiebe_mott_boyer_lester_2020} use cluster algorithms to identify student misconceptions in a block-based programming environment for non-CS major students based on program structures. 
%The clustering results could help researchers identify which students
%would be benefited from certain feedback. 
Wiggins et
al. \cite{wiggins_fahid_emerson_hinckle_smith_boyer_mott_wiebe_lester_2021}
finds five major clusters of hint requests in a block-based
programming system equipped with an intelligent tutor. 
% and investigate how students' perceived computer knowledge and skill were related to their help-seeking behaviours.
Hossein et al. \cite{DBLP:conf/ppig/HosseiniVB14} leverages clustering
analysis to further investigate the correlation between students’
programming speed and programming behaviours by collecting programming
snapshots whenever an action occurs. They then divide students into two
clusters by comparing a student's programming speed % their number of snapshots 
to the median speed. Lahtinen et al. \cite{lahtinen2007categorization}
uses different levels of Bloom’s Taxonomy as features to
identify six distinct student groups 
%: competent, practical, unprepared, theoretical, memorizing, and indifferent groups that
that instructor should be aware of when teaching introductory programming courses. 

In contrast to these existing works, our work considers multi-categorical features involving the grade, total time spent on the assignment, and the number of static errors encountered to identify clusters of students. 
%Hence, our analysis provides a richer picture. 

% They propose four chronotypes: morning, afternoon, napper and evening and the analysis results aligned with chronotypes of typical populations and supported correlations of chronotypes to academic achievement. 

Based on the identified clusters, we follow existing work in understanding the work/rest patterns of students. In particular, Claes et al. \cite{claes_2018} study programmers' working patterns using clustering analysis on time stamps of committed activities of 86 large open-source software projects. Zavgorodniaia et al. \cite{zavgorodniaia_shrestha_leinonen_hellas_edwards_2021} study the chronotypes of students through cluster algorithms using keystroke data. In our study, we use activity data (such as whether a student compiled or graded their homework) to study the work/rest patterns of students. It is the first study in the context of typed functional programming.

We further analyze static errors in typed functional programming
assignments and their impact on different student clusters. This is the first such study in this setting. Previous studies focus 
on compilation events in object-oriented programs written in Java. For example, Ahmadzadeh et al. \cite{ahmadzadeh_elliman_higgins_2005} investigates compiler error frequencies of student programs and debugging activity patterns in Java. They suggest debugging skills should be emphasized in the teaching of programming. Altadmri et al. \cite{altadmri_brown_2015} collect a large dataset comprising compilation events of 250,000 students, which provides a robust analysis of error patterns and time for fixing different errors. Denny et al. \cite{denny_luxton-reilly_tempero_2012} also study various syntax error frequencies and how long students spend fixing common syntax errors. They also found that certain types of errors remain challenging even for higher-ability students. Edwards et al. \cite{edwards_kandru_rajagopal_2017} analyze 10 million static analysis errors found in over 500 thousand program submissions made by students over a five-semester period. The experimental results suggest error frequencies made by CS major and non-major students are consistent.

Our analysis is one of the first that investigates in more depth the
frequency of various static errors in the typed functional programming
assignments.  Here, static errors go beyond syntax and simple type errors
and include for example detection of missing branches in a program. 
% It also shows that our identified student clusters differ significantly in their ability to fix static errors.  

\section{Study Design}
% \subsection{Research Questions and Hypothesis}
% \subsection{Research Questions}
This research aims to gain a deeper understanding of how students
develop typed functional programs (TFP). We assume that the grade
alone is not a good indicator of how well a student masters basic
concepts and achieves competency. Instead, we propose that taking into
account the time spent as well as the number of errors a student
encounters can provide a more nuanced picture. Hence, the main question that we tackle in this paper is how can we best identify different clusters of students taking into account grades, time spent, and the number of errors. We then analyze our clusters with respect to five hypotheses:
\begin{description}
\item[H1:] Even students with a high grade in programming assignments may significantly struggle with a range of static errors.
% -- from syntax and type errors to more conceptual errors such as missing branches in a program.

\item[H2:] Despite a lower grade, students who spend less time and have a low number of static errors do in fact well overall. 

\item[H3:] Work/rest patterns of students as well as the time a student spends on homework play a role in students achieving a high grade. It highlights how driven a student is. % It may however highlight more how driven a student is rather how well they understand basic concepts.

\item[H4:] Static errors in TFP range from syntax and type errors to detecting unbound variables and missing branches in programs. This wide range of static errors provides a fine-grained picture of concepts students find challenging. 

\item[H5:] Error fix ratio, i.e. how many tries a student needs to fix a static error, indicates how well students understand basic ideas in TFP and this is correlated to their understanding and performance.

    % \item \textbf{RQ1:}  What static errors do student experience the most in learning functional programming? 
    % \item \textbf{RQ1:}  How can we identify different types of learners? 
    % \item \textbf{RQ2:}  Can we identify different work / rest patterns (chronotypes) of students? If so, how do chronotypes relate to performance clusters? 
    % \item \textbf{RQ3:}  When do students start and finish their homework?
    % \item \textbf{RQ4:}  What are error patterns of students of different performance clusters? 
\end{description}

\subsection{Course Context}
%  and Online Programming Platform
%\subsubsection{Course Context}
Our study concerns students in a second-year undergraduate computer science course at McGill university. The course introduces concepts about functional programming and programming paradigms. It is offered every semester with more than 300 registered undergraduate students. In this study, all data is collected in the Fall 2021 semester when students could attend online Zoom or in-person sessions. 

\begin{table} %ok
\small
  \begin{tabular}{|l|l|r|r|}
    \hline
     & programming topics
     & \#tasks \\
    \hline
    HW1 & basic expressions, recursion & 7    \\
    HW2  & data types and pattern matching & 6  \\
    HW3  & higher-order functions & 11   \\
    HW4  & references, state, memorization & 5  \\
    HW5  & exception, continuations & 5    \\
    HW6  & lazy programming, toy language & 5   \\
    \hline
  \end{tabular}
  \caption{Overview of six programming assignments. 
%The first column shows the homework number, whereas the second column describes the topics assessed in each assignment. The third column shows the number of programming tasks and test cases required for each assignment.
}
% \textcolor{red}{add time, difficult}
  \label{tb:overview}
\end{table}
The course had six bi-weekly programming assignments each 
worth 5\% of the final grade. 
%The expected time for these assignments was approximately 6h. The bulk of the grade, 70\%,  was determined through an online midterm and an in-person final exam. 
Each homework consists of several programming tasks to implement functions and test cases. Homework information is summarized in Table~\ref{tb:overview}. All homework assignments were hosted on Learn-OCaml~\cite{canou16}, an online programming platform for OCaml which allows students to edit, compile, test, and debug code all in one place. We used a modified version of Learn-OCaml by Hameer and Pientka \cite{aliya} with additional features such as style checking and evaluation of test cases written by students. 
%The former detects stylistic issues and sends hints to students to encourage them to use good programming idioms, while the latter practices students' skills of designing test cases to differentiate \textit{unseen} common buggy implementations from correct reference implementations. 

\subsection{Data Collection }\label{sec:data-collection}
\begin{figure}[t]
    \centerline{\includegraphics[scale=0.155]{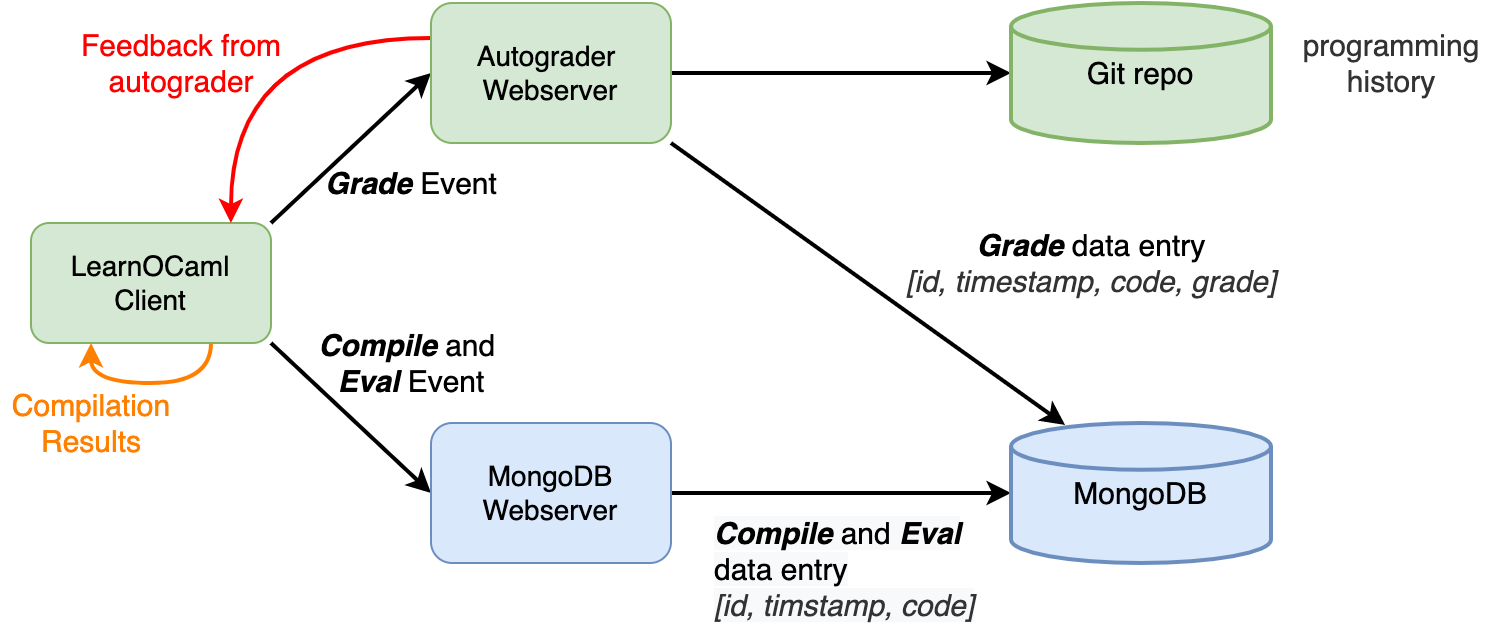}}
    \caption{Data collection pipeline. \textmd{{\tt{Grade}} and {\tt{Compile}} and {\tt{Eval}} events are handled by different servers, all submission data are stored in a \textit{MongoDB} database. The components highlighted in light green are original components in the Learn-OCaml platform, while the components highlighted in light blue are newly introduced by us.} }
    \label{fig:pipeline}
\end{figure}

Our data collection pipeline is built on top of the Learn-OCaml platform and it can automatically log students' actions. Specifically, we send local programming events like compile and evaluation (for testing and debugging) with asynchronous logging requests to our backend server.
Figure~\ref{fig:pipeline} illustrates the process of collecting the data from the online environment Learn-OCaml. 

Around 52.81\% (i.e., 169 out of 320) students gave us consent to access their data. We collect more than 270,000 programming events, and each event stores a snapshot of the code as well as feedback information (e.g., time{-}stamp, static errors, grades, etc.). 
%Such fine-grained data is beneficial for us to understand how students interact with online programming environments with automated graders and feedback support. It can also help us understand students behaviours, such as work habits and common programming mistakes, and cluster students into different categories based on the collected information.

\subsection{Feature}
% \subsubsection{Feature}
For each homework, we collect a sequence of programming \emph{activity
  events}. The activity events include grade, compile, and evaluation events. This allows us to create an \textbf{activity density vector} for each student. It is a four-element vector that represents the percentage of the student’s activity events that occurs in different ranges of hours [0-6, 6-12, 12-18, 18-0], which is the same choice of ranges suggested in \cite{zavgorodniaia_shrestha_leinonen_hellas_edwards_2021}.

%\textbf{Activity density vector}. A four-element vector that represents the percentage of the students activity events that occurs in different ranges of hours [3-9, 9-15, 15-21, 21-3], which is the same choice of ranges suggested in \cite{zavgorodniaia_shrestha_leinonen_hellas_edwards_2021}.

In addition, we design the following features based on the activity event sequence:

\begin{itemize}
\item \textbf{Start time}. The day when a student starts \textit{actively} working on an assignment based on the activity events collected.
% Defined as the first day when a student has at least k programming activity events. We define k = 5 based on experimental results.

\item \textbf{End time}. The day when a student finishes an assignment, which is the last {\tt{Grade}} event. 

\item \textbf{Working session}. Defined as the 
  time window where activity events occur. If there is no activity
  event within 30min, then the working session is assumed to have ended. % window in which a student is active (i.e. an activity event occurs). 

\item \textbf{Total time spent}. Sum over the length of all working sessions.
% The sum of the length of work sessions is a genuine metric of students' time spent on homework, measured in Hours.

\item \textbf{Number of errors}. The number of static errors that a student made while completing an assignment.

\item \textbf{Grade}. The final grade a student receives for an assignment. 
\end{itemize}

\subsection{Feature Engineering}
There are two challenges to applying clustering algorithms and
statistical tests to our study. The first one is skewed data 
. For instance, the grade is highly skewed as students can always improve their grades through interacting with the auto-grader. The second one is the difference between feature scales, which renders the clustering results incoherent. We use two approaches to address these challenges. First, we use non-parametric tests including Spearman correlations and Kruskal-Wallis H-Tests. Second, we apply the rank transformation on features to facilitate clustering algorithms.

\section{Identifying Student Clusters}\label{sec:ans}
To identify student clusters, we run the K-means\cite{hartigan1979algorithm} clustering algorithm on the aggregation (mean) of three most important features (i.e., \emph{grade}, \emph{number of errors} and \emph{time spent}) over six homework. We use the \textit{elbow method} to determine the optimal k (the number of clusters) to be 4. After determining the optimal k, we re-run the K-means algorithm and report the results in Table \ref{cls_table}. We give the time in hours and note that all clusters have a similar size in terms of number of students ($\#Std$).

\newcommand{\stdev}[1]{\footnotesize{($\pm$ #1)}}
\begin{table}[b]
\small
  \begin{tabular}{|l|r@{~}|c|r|r|}
    \hline
    \multicolumn{1}{|c|}{Clusters} & $\!\!$\#Std & Time (Hours) & \multicolumn{1}{c|}{\# Error} & \multicolumn{1}{c|}{ Grade} \\
    \hline
     {A - Quick learning}$\!\!$ & 46 &  5.30 \stdev{0.94} & 66.11  \stdev{26.95} & 95.24 \stdev{3.25}   \\ \hline
     {B - Hardworking}     & 46 &  8.24 \stdev{1.52} & 148.67 \stdev{63.26} & 94.25 \stdev{3.90} \\ \hline
     {C - Satisficing}    & 31 &  4.47 \stdev{1.01} & 52.26  \stdev{21.89} & 74.43 \stdev{11.31}   \\ \hline
     {D - Struggling}     & 46 &  6.49 \stdev{0.94} & 118.14 \stdev{35.32} & 72.81 \stdev{11.03} \\ \hline
  \end{tabular}
\caption{Student clusters}

\label{cls_table}
\end{table}

To determine whether the resulting four clusters are different, we run
a Kruskal-Wallis H-Test, which is a nonparametric equivalent of an
ANOVA, on the three features (time spent, \#errors, and grade) of each
cluster. The results are statistically significant with the
statistics of 113.26, 100.87, and 123.02 respectively, and all p-values < 0.0001. This suggests the four clusters are statistically different.

Students in cluster A have the highest average grade (95.24) while spending less than the expected 6h on solving the homework. 
% the second-lowest average time spent (5.30) and the second highest number of errors (66.11).
 This suggests that they achieve their goal with relative ease. In fact, students in this cluster outperform students in other clusters by a large margin. We characterize this cluster as \textbf{quick learning}. 

Students in cluster B have the second-highest average grade (94.25). However, they also have the highest average
number of errors (148.67) and with 8.24h spend significantly more time on homework than any other group. In particular, they spend significantly more time than expected. This suggests that they face many difficulties which they manage to overcome by spending a significant amount of time. These students are driven to improve their work and to achieve the highest possible grade. Hence, we characterize them as \textbf{hardworking}. This data supports our hypothesis \textbf{H1}.

Cluster C has the lowest average number of errors (52.26) and spent the least amount of time (4.47h) on the homework. With an average grade of 74.43, they still achieve a ``good enough'' result. These students achieve their goals by saving time and effort. %This is in fact a very rational choice given that loosing 1\% or less on a given homework assignment may have little impact on the overall grade.
 At the same time, these students reach a satisfying level of competency as evidenced by their low number of average errors. We describe these students as \textbf{satisficing} students. This supports our hypothesis \textbf{H2}.

Students in Cluster D are in fact closely related to students in cluster B, which shows a similarly high average number of errors (118.14) and a significant amount of time (6.49h). However, compared to students in cluster B, they fail to overcome the difficulties along their path. These students are \textbf{struggling}.

\section{Understanding student clusters}
\subsection{How do work habits vary for different student clusters?}
To investigate our hypothesis \textbf{H3}, we consider when students are active based on our activity data. 
Prior research suggests that chronotype, a person's preference in carrying out activity at certain periods in a day, 
 is governed by the circadian cycle which is controlled by clock genes \cite{reppert_weaver_2002,dibner_schibler_albrecht_2010}. In this section, we are interested in investigating the chronotypes, or in other words, the work habits of students. 
In particular, it has been observed that  ``morningness'' is positively correlated with academic achievement \cite{zavgorodniaia_shrestha_leinonen_hellas_edwards_2021,preckel2011chronotype}.

To identify potential chronotypes, we run the K-means clustering algorithm on the feature space spanned by activity density vectors. The elbow method yields $k=3$, suggesting three possible chronotypes, which is different from four chronotypes reported in \cite{zavgorodniaia_shrestha_leinonen_hellas_edwards_2021}.  We report centroids of each chronotype cluster in Table \ref{table_chrono}. 

\begin{table}[bht] %ok
\small 
\begin{tabular}{ |l|rrrrc|} 
\hline
  Chrono clusters & \ 0 - 6 & \ 6 - 12 & \ 12 - 18 &\ 18 - 0 &\ Chronotype  \\
 \hline
Cluster 1 & 8\%  & 14\% & 26\%  & \bf{52\%} & Evening (Eve)\\ 
Cluster 2 & 4\%  & \bf{26\%} & 20\% & 50\% & Morning (Mor)\\ 
Cluster 3 & 2\% & 19\%	& \bf{37\%} & 42\% & Afternoon (Aft) \\ 
 \hline
%  \hline
\end{tabular}
\caption{Centroids of each chronotype.}
\label{table_chrono}
\end{table}

As we can see, most activities occur from 18:00 - 00:00 for all three clusters. This is not surprising as most students may have classes during the day. Based on this observation, we aim to define chronotypes by considering secondary activity peaks as well. We notice that Cluster 2 has its secondary activity peak (26\%) in 6:00 - 12:00 whereas Cluster 3 has the secondary activity peak (37\%) in 12:00 - 18:00. Thus, we define Cluster 2 and 3 as the \textbf{morning} (Mor) and \textbf{afternoon} (Aft) type. Cluster 1 has only one activity peak in 18:00 - 00:00, thus we define it as \textbf{evening} (Eve) type.

% \begin{table}[htbp]
% \small 
% \begin{tabular}{ |l|rrrr|} 
% \hline
% Clusters   & Morning & \ Afternoon &\ Evening \\
%  \hline
%  Cluster1 	  & 148.67 & 108.39	 	& 108.39 \\ 
%  Cluster2  & 118.14 & 83.73	  & 108.39 \\ 
%  Cluster3  & 52.26	 & 85.60   & 108.39 \\ 
%  Cluster4 	 & 66.11 & 	109.53   & 108.39 \\ 
%  \hline
% \end{tabular}
% \caption{The four performance clusters of students.}
% \label{cls_table}
% \end{table}

\begin{table*} %ing
\small 
\begin{tabular}{ |l|l|rrrrrr|} 
\toprule
% \hrule
 Error Groups & Error Categories & \ HW1 & \ HW2 & \ HW2 &  HW4 &  \ HW5 & \ HW6\\
\hline

\multirow{2}{*}{\textbf{A. General Static Errors}} 
 &1. Type Error &  38.12\% & 30.94\%  &40.93\%  & 32.65\% & 36.90\% & 34.83\% \\ 
 & 2. Syntax Error &  42.33\% & 21.54\%  &21.79\%  & 32.68\% & 17.80\% & 25.66\% \\ 
\hline
 &3. Unbound value &10.42\% & 7.19\%  & 9.06\%  & 13.42\% & 7.02\% & 7.27\% \\
  \multirow{3}{*}{ \textbf{B. Imperative Thinking Errors} }
  &4. Missing else branch & 1.92\% & 0.75\%  &0.43\%  &0.08\% & 1.03\% & 1.07\% \\
 &5. Unused variable & 0.74\% & 0.65\%  &0.63\%  & 6.37\% & 21.34\% & 7.23\% \\
 \hline
\multirow{2}{*}{ \textbf{C. Pattern Matching Errors}} 
 &6. Pattern matching type error & 0.84\% & 5.24\%  &2.13\%  & 0.62\% & 1.37\% & 1.40\% \\
 &7. Non-exhaustive pattern matching  & 1.02\% & 16.78\%  & 15.74 \%  & 2.47\% & 4.62\% & 11.92\% \\ 
 \hline
\multirow{2}{*}{ \textbf{D. Function Applications Errors}}
&8. Wrong number of arguments& 1.67\% & 2.19\%  &3.38\%  & 1.17\% & 2.09\% & 1.89\% \\
&9. Misuse of non-function values &2.50\% & 2.10\%  &2.07\%  & 1.72\% & 1.50\% & 1.77\% \\ 
\hline
&10. Others & 0.88\% & 12.6\%  &5.89\%  & 8.83\% & 6.33\% & 6.96\% \\ 
% \bottomrule
\hline
& Total number of errors  & 7,850 & 27,519 & 14,331 & 19,859 & 22,467 &26,681\\
 \hline
\end{tabular}
\caption{Error Groups and error categories together with their distribution of HWs}
\label{table-errors}
\end{table*}

\begin{figure}[htbp]
    \centerline{\includegraphics[scale=0.35]{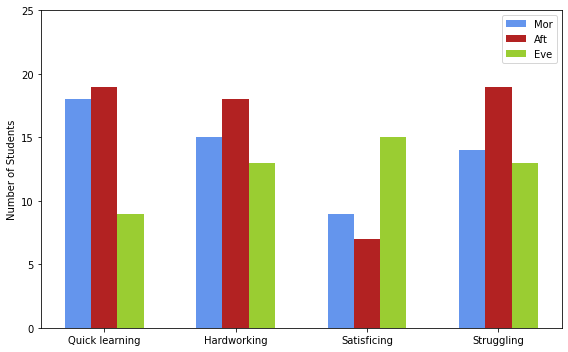}}
    \caption{Chronotype distribution in each student cluster.}
    \label{fig:chrono_cls}
\end{figure}

As Figure \ref{fig:chrono_cls} suggests, quick-learning students usually tend to work in the morning and afternoon whereas satisficing students worked on their homework in the evening. This suggests quick-learning students were driven, motivated, and had possibly better time management skills. In general, satisficing students were the only group to have a strong incline to work in the evening. This could point to other commitments that students have or a high course load.
% This may have many reasons, yet it highlights a choice made by these students.
The afternoon type occurs most frequently in struggling and hardworking clusters. This may be because they were seeking help during office hours that were offered during the day or they simply required more time in general. Overall, our results confirm previous findings that certain chronotypes are related to academic achievement\cite{zavgorodniaia_shrestha_leinonen_hellas_edwards_2021,preckel2011chronotype}.

%labels = ['Hardworking', 'Struggling', %'Satisficing', 'Quick learning']
\begin{figure}[htbp]
    \centerline{\includegraphics[scale=0.235]{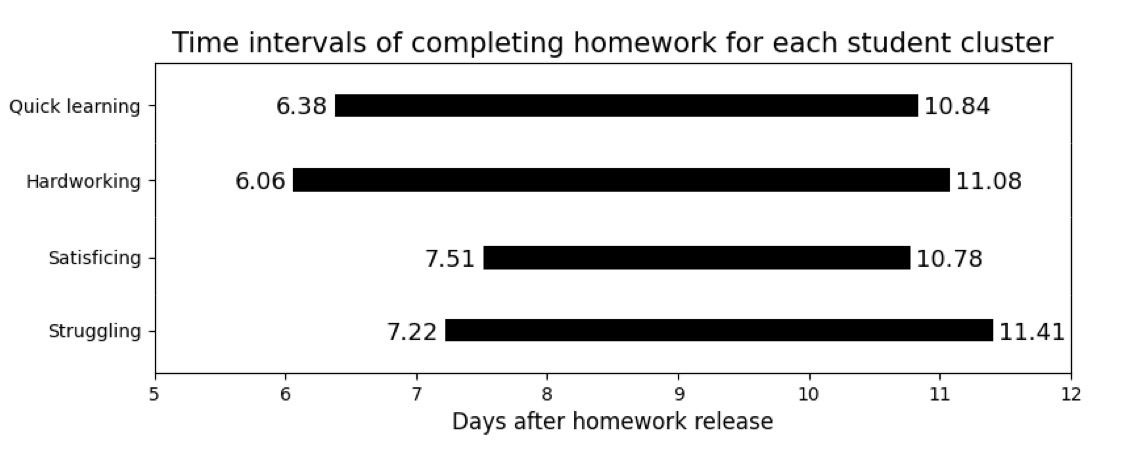}}
    \caption{Clustering result of different types of students \textmd{The start of a time interval stands for the average start time whereas the end represents the average end time.}}
    \label{fig:interval}
\end{figure}

\subsection{How long do different clusters of students work on their homework?}
To further investigate hypothesis \textbf{H3}, we investigate when students in a given cluster start and finish their homework. We report the average start time and end time for each cluster in Figure \ref{fig:interval}. In addition, the Kruskal-Wallis H-Test suggests start date was statistically significantly different (\emph{stat} = 22.59, \emph{p-value} < 0.0001) whereas the end date was not (\emph{stat} = 3.12, \emph{p-value} = 0.37). Despite that, we can still observe some interesting patterns. 

We note that both satisficing and struggling students start relatively late on their homework, at 7.51 and 7.22 average days respectively. However, satisficing students finish the earliest (10.78). This underscores the fact that they accept a ``good enough'' result rather than striving for better outcomes.  Further, satisficing students had the shortest working duration. This substantiates our claim that these students achieve their goals by saving time and effort.

Struggling students experienced many difficulties as evidenced by a high number of static errors that they encounter. These students finish indeed last (finish time (11.41)). This indicates that these students are struggling, although they do try their best until the very end. However, they lack the skills or support to overcome their difficulties. 

Hardworking students have the longest time interval. While they start the earliest (6.06), they finish the second latest (11.08). This shows the commitment and dedication they bring to their work. 
%However, we suspect that they are aware of their difficulties which may be one of the reasons for starting early.

Quick-learning students tend to start quite earlier (6.38), although not as early as hardworking students. This suggests that these students have confidence in their abilities to finish the homework smoothly.

 We ran Spearman correlations to examine the correlation between start time and homework grade, the statistically significant result (\emph{correlation} = -0.42, \emph{p-value} < 0.0001) suggests procrastination affects negatively on student learning outcomes, which has been widely reported \cite{hussain_sultan_2010,hailikari_katajavuori_asikainen_2021,pmid:24644450}.

% Using normality tests, we found that the data was not normally
% distributed and therefore we applied non-parametric tests for our
% analysis. This includes using Mann-Whitney U tests instead of independent sample t-tests, Spearman correlations instead of Pearson
% correlations, and Kruskal-Wallis H-Tests instead of ANOVAs. 

% How do static errors affect students in different clusters?
\subsection{How do static errors affect students in different clusters?}
Compilers for typed functional programming languages such as OCaml provide a wealth of errors and feedback to programmers. It not only reports syntax and type errors but also reports, for example, unused variables, and missing branches in case-statements and if-expressions. This provides a basis for a better understanding of what basic concepts students struggle with the most.

\subsubsection{Overview of static errors}\label{subsec:static-error}
To investigate our hypothesis \textbf{H4}, we analyze the types of errors of each failed compile event and group errors into four main categories: general static errors (eg. group A), errors due to imperative thinking (Group B), and errors related to pattern matching and function (eg. groups C and D).  We also include how often particular errors occurred in assignment submissions (see Table~\ref{table-errors}). 

The first homework shows a significant spike (42.33\%) in syntax errors encountered. 
%the first homework includes ``fix-me'' exercises where students need to fix syntax and type errors in a given program. 
This is unsurprising, as it is the first time that students attempt to write programs in a new language. However, it may be surprising that 20\% to 30\% of the errors encountered are related to syntax and type errors (Group A) throughout the semester. In fact, these errors constitute around 60\% of errors for every homework assignment in Table~\ref{table-errors}. This may point to the fact that type errors in TFP catch conceptual errors in the programmer's thinking early rather than later during testing. 
%However, it seems surprising that type errors play such a significant role. 
This may also suggest instructors dedicating more time to demystifying type error analysis. 
%Presently, this knowledge is taught more implicitly  along with other concepts such as higher-order functions or pattern matching.

For some key concepts from typed functional programming such as pattern matching, our error analysis indicates that students do improve and gain a better understanding of it. When pattern matching is first introduced in HW2, pattern matching errors and non-exhaustive pattern matching errors (Group C) consist 22\% of total static errors. After practicing HW2 and HW3, the proportion of Error Group C drops greatly, which suggests that students gain a deeper understanding with more programming practice.

 One of the prerequisites of this course is taking an introductory CS course, which is taught in Java or Python at our university. This implies that all of the participants had experience in programming before and had to deal with conceptual transfer from imperative/object-oriented programming (Python or Java) to functional programming (OCaml). 
%As OCaml is a statically typed programming language, it is expected that many students experienced great difficulties in getting familiar with syntax and type inference mechanisms. Therefore, 
Students usually report transitioning smoothly between procedural language and object-oriented language for concepts such as \textit{if-conditionals} and \textit{functions and scope}\cite{10.1145/3372782.3406270}. From our observations, students struggle more when transitioning to functional programming. In particular, they struggle with the concept of bound or unbound variables, missing branches in if-expressions, and function application errors. Although these errors occur less frequently than syntax and type errors, we believe it highlights that students struggle with thinking recursively and considering all cases in such a recursive program (Error No.4,7). Therefore, if-else expression without an else branch also often leads to type errors in a language like OCaml. % Additionally, pattern matching is specifically used in functional programming to define functions by recursive case analysis. If students fail to handle recursive cases thoroughly, non-exhaustive pattern matching errors and missing else branch errors will emerge often.

Moreover, imperative programming supports variables declared in the local or global state, while in functional languages, such as OCaml, we distinguish between stateful variables that can be updated and bound variables. While the concept of free variables and bound variables and the difference between stateful variables are discussed frequently in this course, students continue to encounter errors related to variables. In particular, the unbound value error occurs throughout the semester. This seems to be a sign that the concept of stateful variable declarations as used in imperative programming is persisting in how students think about a given problem. The most essential concept of functional programming is that functions are first-class citizens. Therefore, higher-order functions, which take a function as an argument, or return a function, are used frequently, especially in HW3 and subsequent assignments. If functions are not used correctly, it would most frequently be flagged as a type error. However, OCaml also provides other error reporting. In particular, it may report on the incorrect number of arguments (Error NO.8) and use a function value instead of applying arguments on a non-function value (Error NO.9). These errors form a non-negligible class indicating where students stumble.

\subsubsection{ How efficiently do students in each cluster fix errors?}

Lastly, we investigate hypothesis \textbf{H5} and aim to understand how students in different clusters vary in their ability to fix errors quickly. Table~\ref{tb:succ-fail} shows the average number of successful compile events and failure ones experienced by different student clusters throughout the semester. The Failure/Success ratio \textit{x} can be roughly interpreted as \textit{debugging efficiency} or \textit{error fix rate} that it on average costs a student \textit{x} failure compile events to get a successful one. 

\begin{table}[htbp] %ok
\small
  \begin{tabular}{|l|rrrr|}
    \hline
     &  Quick-learning   & Hardworking    & Satisficing & Struggling    \\
    \hline
    Success  & 37.9 & 60.4 & 28.1 & 40.7 \\
    Failure  & 85.7 & 162.3 & 66.9 & 118 \\
    F/S      & 2.26 & 2.67 &  2.38 & \textbf{2.90} \\
    \hline
  \end{tabular}
  \caption{Average success, failure and failure/success ratio (F/S) of compile events in each student cluster}
  \label{tb:succ-fail}
\end{table}

Struggling students have the most difficulty in fixing static errors, requiring 2.9 failure compilations to fix the error on average. By contrast, quick-learning students have the best ability to debug with only a 2.26 failure compilation to get a successful one. 
% \textcolor{red}{We run KMann–Whitney U test to determine whether quick learning students has a smaller Failure/Success ratio than struggling students. The result is statistically significant with the \emph{statistic} of 8.79, and \emph{p-value} < 0.05.}
Furthermore, the gap between their debugging efficiency is more significant, if we look at their average failure and success. While the average success for struggling students (40.7) and quick learners (37.9 ) are close, their average failures have a substantial gap: a struggling student has around 30 more failure compilations than quick learners. 

\begin{figure}[htbp]
    \centerline{\includegraphics[scale=0.35]{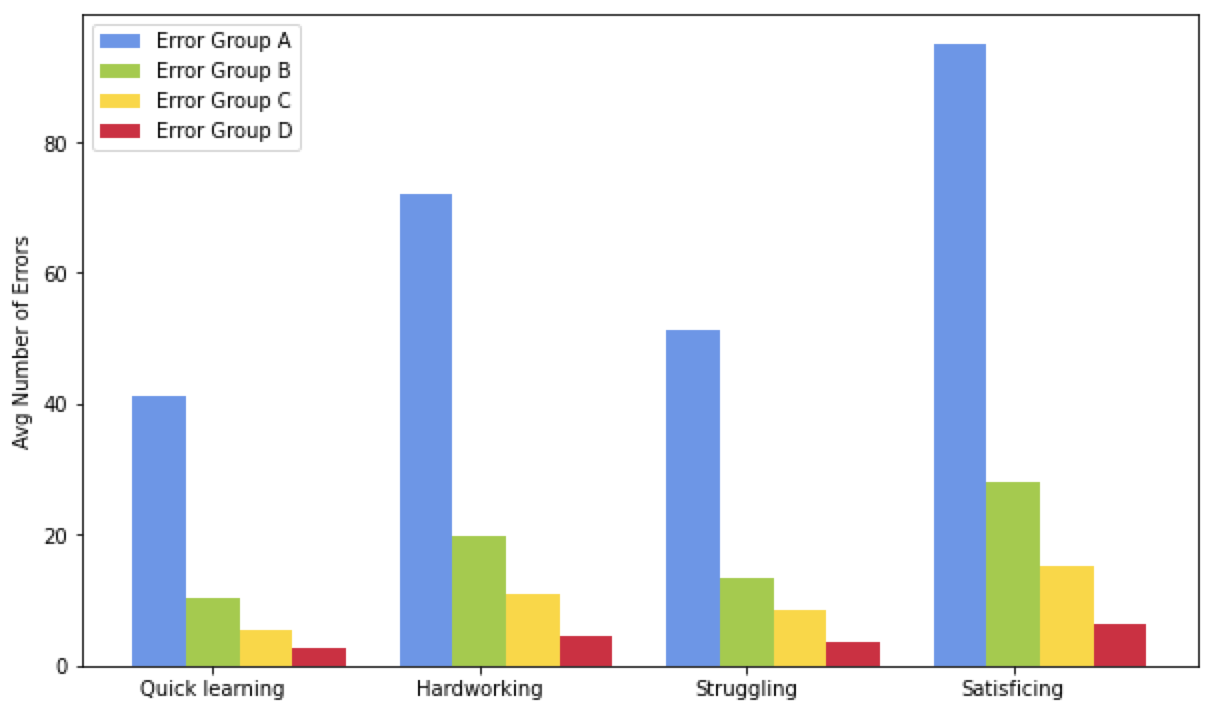}}
    \caption{Distribution of static errors in each student cluster.}
    \label{fig:error-dist}
\end{figure}

The row of Failure in Table \ref{tb:succ-fail} can be further represented by the average number of each group of static errors for four student clusters in Figure \ref{fig:error-dist}.
Type and syntax errors (Group A) dominate for all clusters but there are noteworthy differences.
Quick learners have fewer errors in all groups, not only general static errors but also errors specific to functional programming. Satisficing students have the fewest errors in Group B, C, and D which may indicate that they in fact achieve competency. Lastly, hardworking and struggling students have significantly more errors in all error groups. In particular, they struggle more with basic concepts such as bound or unused variables, missing branches, and the proper use of functions.

\section{Conclusion}\label{sec:conclusion}
\ 
\indent 
In this study, we aim to understand how students develop functional programming assignments based on data collected through the Learn-OCaml programming platform. Our analysis considers grade, total time spent, and the total number of static errors to identify four student clusters:  \emph{"Quick-learning"}, \emph{"Hardworking"}, \emph{"Satisficing"}, and \emph{"Struggling"}.  Using statistical tests we validate our clustering results along with other analysis results. This provides a nuanced picture of students' behaviours and also exposes different paths towards achieving academic success in the course.  Our analysis of chronotypes confirms that students who work in the morning reach the highest grade most quickly and smoothly. The total amount of time students spend on the homework also highlights the difference and similarities between the different student clusters. Although this part of the analysis was done in the context of a functional programming course, we expect our methodology to be applicable to other programming courses and help identify clusters of students who would benefit from additional support. 

Our detailed analysis of static errors in typed functional programming also highlights areas where instructors can adjust their course content and possibly revisit topics. We believe our analysis also provides insights for students themselves, in particular the hardworking students, to understand which aspects they still struggle with and to seek clarifications. This would possibly allow them to become more efficient debuggers, spend less time on homework assignments, and improve their conceptual understanding. 

% for assessing students’ behaviours and performance. Our structured analysis covers three important topics of learning analytics: static error analysis, student behaviors clustering, and work/rest pattern analysis. To be more specific, we identify four performance clusters of students - using clustering algorithm. 

% Although our analysis results may not generalize well to other courses and learning experience, we believe our features and analysis methodology are meaningful and they could be applicable to not only some traditional classes utilizing online learning and grading system but also MOOC, as they both possess valuable data regarding students' work habits and performance. This would improve the learning experience since the instructors would precisely know the challenges encountered by their students.

%%
%% The acknowledgments section is defined using the "acks" environment
%% (and NOT an unnumbered section). This ensures the proper
%% identification of the section in the article metadata, and the
%% consistent spelling of the heading.
% \begin{acks}
% To Robert, for the bagels and explaining CMYK and color spaces.
% \end{acks}

%%
%% The next two lines define the bibliography style to be used, and
%% the bibliography file.
\bibliographystyle{ACM-Reference-Format}
\balance
\bibliography{main}

%%
%% If your work has an appendix, this is the place to put it.
\appendix

\end{document}